\documentclass{revtex4}

\usepackage{graphicx}
\newcommand{\LL}{\mathrm{LL}}
\newcommand{\LR}{\mathrm{LR}}
\newcommand{\RL}{\mathrm{RL}}
\newcommand{\RR}{\mathrm{RR}}

\newcommand{\eemm}{\mathrm{e}^+\mathrm{e}^- \rightarrow \mu^+ \mu^-}

\newcommand{\bb}{\mathrm{b \bar{b}}}

\newcommand{\be}{\begin{equation}}
\newcommand{\eq}{\end{equation}}
\newcommand{\ba}{\begin{eqnarray}}
\newcommand{\ea}{\end{eqnarray}}

\begin{document}
\bibliographystyle{revtex}

\preprint{ }

\title{Probing New Scales at a $e^+e^-$ Linear Collider}



\author{Marco Battaglia}
\email[]{Marco.Battaglia@cern.ch}
\affiliation{CERN, CH-1211 Geneva 23, Switzerland}

\author{Stefania De Curtis}
\email[]{decurtis@fi.infn.it}
\author{Daniele Dominici}
\email[]{dominici@fi.infn.it}
\affiliation{Dip. di Fisica, Universita' degli Studi, Firenze, Italy}
\affiliation{INFN, Sezione di Firenze, Firenze, Italy}
\author{Sabine Riemann}
\email[]{riemann@ifh.de}
\affiliation{DESY Zeuthen, Germany}

\date{\today}

\begin{abstract}
\vspace*{0.25cm}
Extending the sensitivity to New Physics beyond the anticipated reach of the {\sc Lhc} 
is a prime aim of future colliders. This paper summarises the potential of an $e^+e^-$ 
linear collider, at and beyond 1~TeV, using a realistic simulation of the detector 
response and the accelerator induced background. The possible LC
energy-luminosity trade-offs offered in probing multi-TeV scales for new phenomena with 
electro-weak observables are also discussed.
\end{abstract}

\maketitle

\section{Introduction}

The {\sc Lhc} is expected to directly probe possible New Physics beyond the Standard 
Model (SM) up to a scale of a few TeV. While its data should provide answers to 
several of the major open questions in the present picture of elementary particle 
physics, it is important to start examining how this sensitivity can be further extended 
at a next generation of colliders.
Today we have a number of indications that New Physics could be of supersymmetric 
nature. If this is the case, the {\sc Lhc} will have a variety of signals to discover 
these new particles and the linear collider (LC) will be required to complement the probe of 
the SUSY spectrum with detailed measurements. However, beyond Supersymmetry there is a 
wide range of other scenarios invoking new phenomena at, and beyond, the TeV scale. 
They are 
aimed at explaininig the origin of electro-weak symmetry breaking, if there is no light 
elementary Higgs boson, at stabilising the SM, if SUSY is not realised in nature, or at 
embedding the SM in a theory of grand unification.

Many of such scenarios predict the existence of new particles that would be manifested as
rather spectacular resonances in $e^+e^-$ collisions, if the achievable centre-of-mass 
energy is sufficient. A high energy LC represents an ideal laboratory for studying this
New Physics~\cite{zp1,dominici}. 
It also retains an indirect sensitivity, through a precision study of the virtual 
corrections to electro-weak observables, when their mass exceeds the available 
centre-of-mass energy. 
This paper summarises the results of a series of studies aimed at quantifying the 
potential of a high energy, high luminosity $e^+e^-$ LC  in extending 
to high scales the probe for New Physics. While a significant activity has already 
addressed the issues related to a TeV-class collider, we now review the potential of 
a multi-TeV LC, such as {\sc Clic}.

\section{Electro-weak Observables at a Multi-TeV LC}

The analysis of the {\sc Lep} and {\sc Slc} data has provided a significant experience 
in the extraction of electro-weak observables, optimising their statistical sensitivity 
and controlling their systematic uncertainties. At larger centre-of-mass 
energies, the relevant $e^+e^- \rightarrow f \bar f$ cross sections are significantly 
reduced and the experimental conditions at the interaction region need to be taken into 
account in validating the anticipated accuracies on the cross section 
$\sigma_{f \bar f}$, forward-backward asymmetries $A_{FB}^{f \bar f}$ and left-right 
asymmetries $A_{LR}^{f \bar f}$ determination at $\sqrt{s}$ = 1~TeV - 5~TeV. Since the 
two-fermion cross section is of the order of only 10~fb, it is imperative to achieve 
high luminosity by reducing the beam-spot sizes. In this regime the beam-beam effects 
are important and the primary $e^+e^-$ collision is accompained by several 
$\gamma \gamma \rightarrow {\mathrm{hadrons}}$ interactions. Being mostly confined in 
the forward regions, this $\gamma \gamma$ background reduces the polar angle acceptance
for quark flavour tagging and dilutes the jet charge separation using jet charge 
techniques. These experimental conditions require efficient and robust algorithms to 
ensure sensitivity to flavour-specific $f \bar f$ production.
The statistical accuracy for the determination of 
$\sigma_{f \bar f}$, $A_{FB}^{f \bar f}$ and $A_{LR}^{f \bar f}$ has been 
studied, for $\mu^+\mu^-$ and $b \bar b$, taking the {\sc Clic} parameters at
$\sqrt{s}$ = 3~TeV. The {\sc Simdet} parametrised detector simulation has been used and
the $\gamma \gamma \rightarrow {\mathrm{hadrons}}$ background, corresponding to 10 
overlayed bunch crossings, has been added to  
$e^+e^- \rightarrow \mu^+ \mu^-$, $b \bar b$ events. $b \bar b$ final states have been 
identified using an algorithm based on the sampling of the decay charged multiplicity of 
the highly boosted $b$ hadrons at {\sc Clic} energies~\cite{bmult}. Similarly to 
{\sc Lep} analyses, the forward-backward asymmetry has been extracted from a fit to the 
flow of the jet charge $Q^{jet}$ defined as 
$Q^{jet} = \frac{\sum_i q_i |p_i T|^k}{\sum_i |p_i T|^k}$, where $q_i$ is the particle 
charge, $p_i$ its momentum, $T$ the jet thrust axis and the sum is extended to all the 
particles in a given jet. Here the presence of additional particles, from the $\gamma 
\gamma$ background, causes a broadening of the $Q^{jet}$ distribution and thus a 
dilution of the quark charge separation. The track selection and the value of the power 
parameter $k$ needed to be optimised as a function of the number of overlayed bunch 
crossings. The results are summarised in terms of the relative statistical accuracies 
$\delta {\cal{O}}/{\cal{O}}$ in Table~\ref{tab:res}. Another important issue is the 
accuracy on the luminosity determination, that needs to be controlled to 0.5\%, or 
better.

\begin{table}
\caption{Relative statistical accuracies on electro-weak observables, obtained for 
1~ab$^{-1}$ of {\sc Clic} data at $\sqrt{s}$ = 3~TeV, including the effect of 
$\gamma \gamma \rightarrow {\mathrm{hadrons}}$ background.}  
\begin{tabular}{|l|c|}
\hline
Observable & Relative Stat. Accuracy \\
           & $\delta {\cal{O}}/{\cal{O}}$ for 1~ab$^{-1}$ \\
\hline \hline
$\sigma_{\mu^+\mu^-}$ & $\pm 0.010$ \\
$\sigma_{b \bar b}$ & $\pm 0.012$ \\
$A_{FB}^{\mu\mu}$ & $\pm 0.018$ \\
$A_{FB}^{bb}$ & $\pm 0.055$ \\ \hline
\end{tabular}
\label{tab:res}
\end{table}

\section{Sensitivity to a $Z'$ Boson}

There is a wide range of New Physics scenarios predicting the existence of new
 vector 
particles with masses in the TeV range.
One of the simplest extensions of the SM consists in the introduction
of  an additional 
$U(1)$ gauge symmetry, whose breaking scale is close to the Fermi scale. 
This extra symmetry is for instance 
predicted in some grand unified theories. 
The extra $Z'$ associated to this symmetry naturally mixes 
with the SM $Z^0$. The mixing angle is strongly constrained by precision 
electroweak data to be of the order of few mrad while direct searches 
at Tevatron for a new $Z'$ boson set a lower mass limit around 600~GeV. Further 
improvements are expected by precision electro-weak data obtained at a {\sc Giga-Z} 
facility.

The search for such an extended gauge  sector offers an interesting framework 
for studying the sensitivity to scales beyond those directly accessible, even at a 
multi-TeV collider. It also raises the issue 
of the ability to discriminate between different models. 
Main classes of
 models with  additional  $Z'$ bosons include  $E_6$ inspired models
and
the left-right models (LR) (for a recent review see \cite{leike}).
Concerning the $E_6$ models, the $Z'$ fermion couplings depend on the angle
$\theta_6$ which defines the embedding of the extra $U(1)$ in the $E_6$ group.

In this study we have considered the so-called 
$\eta$ model with $\theta_6=-\arctan\sqrt{5/3}$, 
the $\chi$ model with $\theta_6=0$
and, as a reference model, the so called
sequential SM (SSM) which has an additional $Z'$ boson 
with SM-like couplings.

There has been a significant interest in the {\sc Lhc} and LC potential in the search 
for a new $Z'$ boson. At the LC, the indirect sensitivity to its mass, 
$M_{Z'}$, can be parametrised in terms of the available integrated luminosity 
${\cal{L}}$, and centre-of-mass energy, $\sqrt{s}$. In fact the 
scaling law for large $M_{Z'}$ can be obtained by considering the effect of the 
$Z'-\gamma$ interference in the cross section $\sigma$. 
For $s<< M_{Z'}^2$ and assuming that the uncertainties $\delta \sigma$ are statistically 
dominated, we get the range of mass values giving a significant difference:
\begin{equation}
\frac{|\sigma^{SM} - \sigma^{SM+Z'}|}{\delta \sigma} \propto \frac{1}{M^2_{Z'}}\sqrt{sL} 
> \sqrt{\Delta \chi^2}
\end{equation}
and the sensitivity to the $Z'$ mass scales as:
\begin{equation}
M_{Z'} \propto (s L)^{1/4}
\label{resc}
\end{equation}

This relationship shows that there is a direct trade-off possible between $\sqrt{s}$ and 
${ \cal{L}}$, which should be taken into account when optimising the parameters 
of a high energy LC. 

\begin{figure}
\begin{center}
\begin{tabular}{l c r}
\includegraphics[width=0.32\textwidth,height=0.4\textwidth]{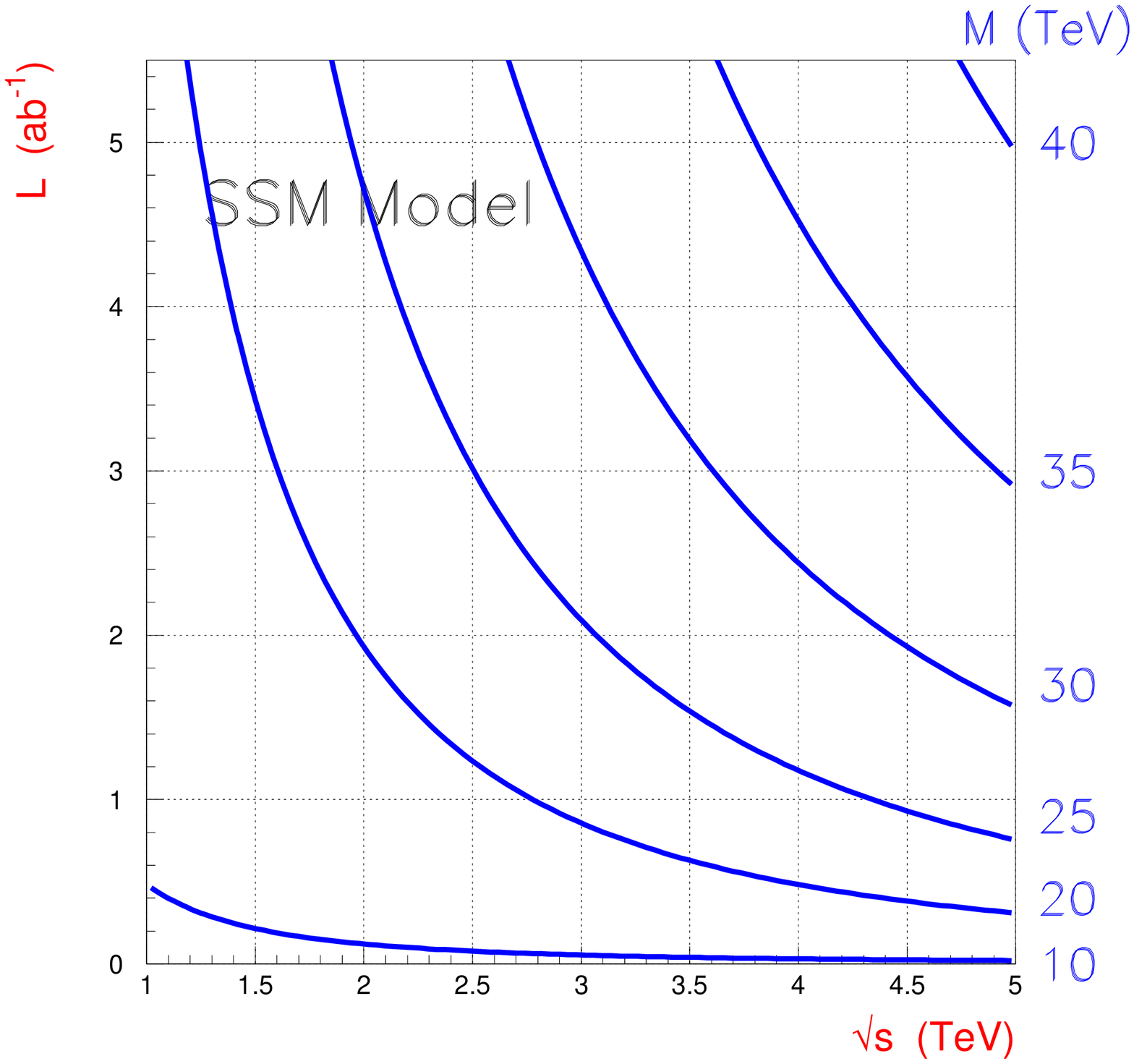}&
\includegraphics[width=0.32\textwidth,height=0.4\textwidth]{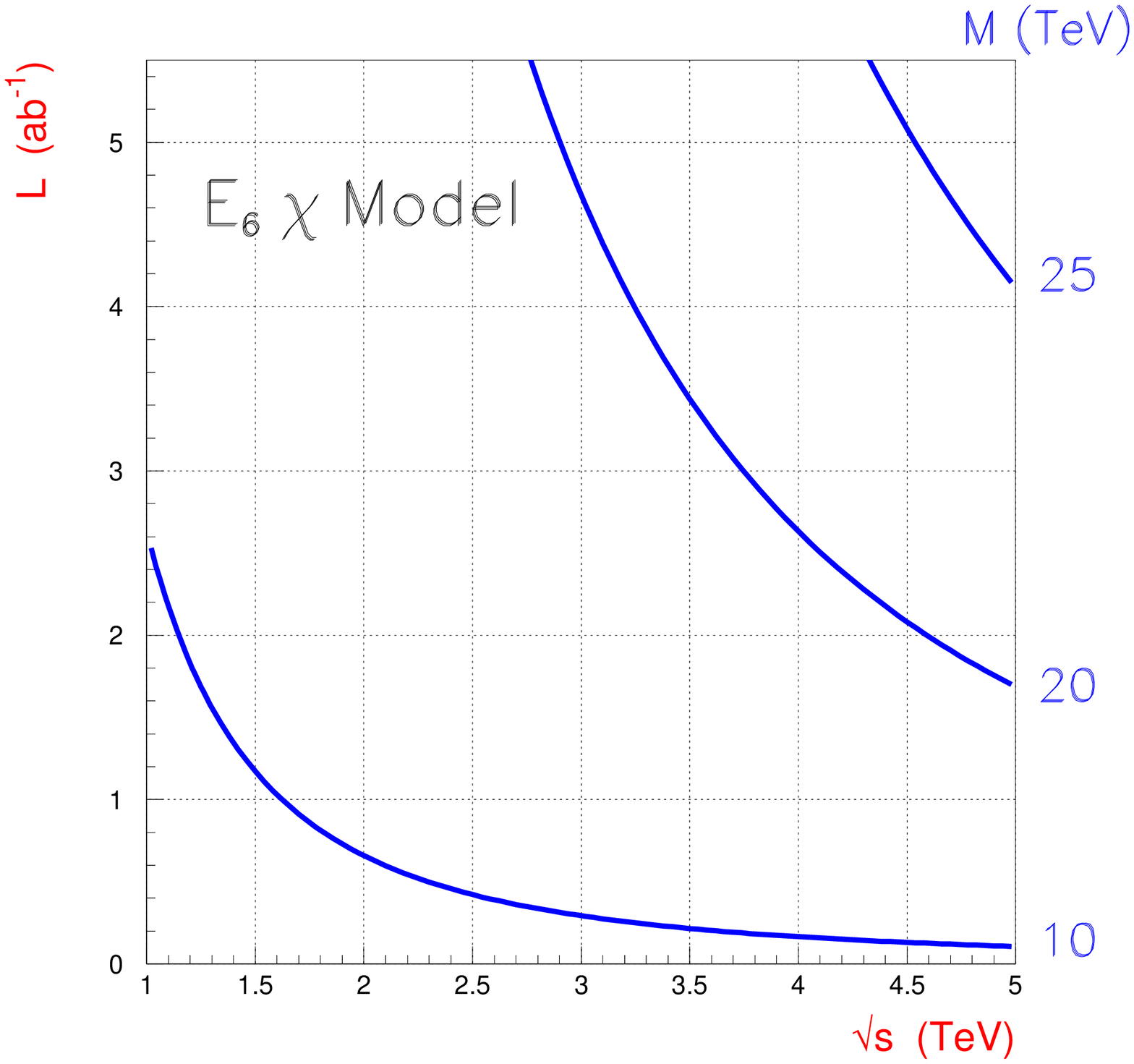}&
\includegraphics[width=0.32\textwidth,height=0.4\textwidth]{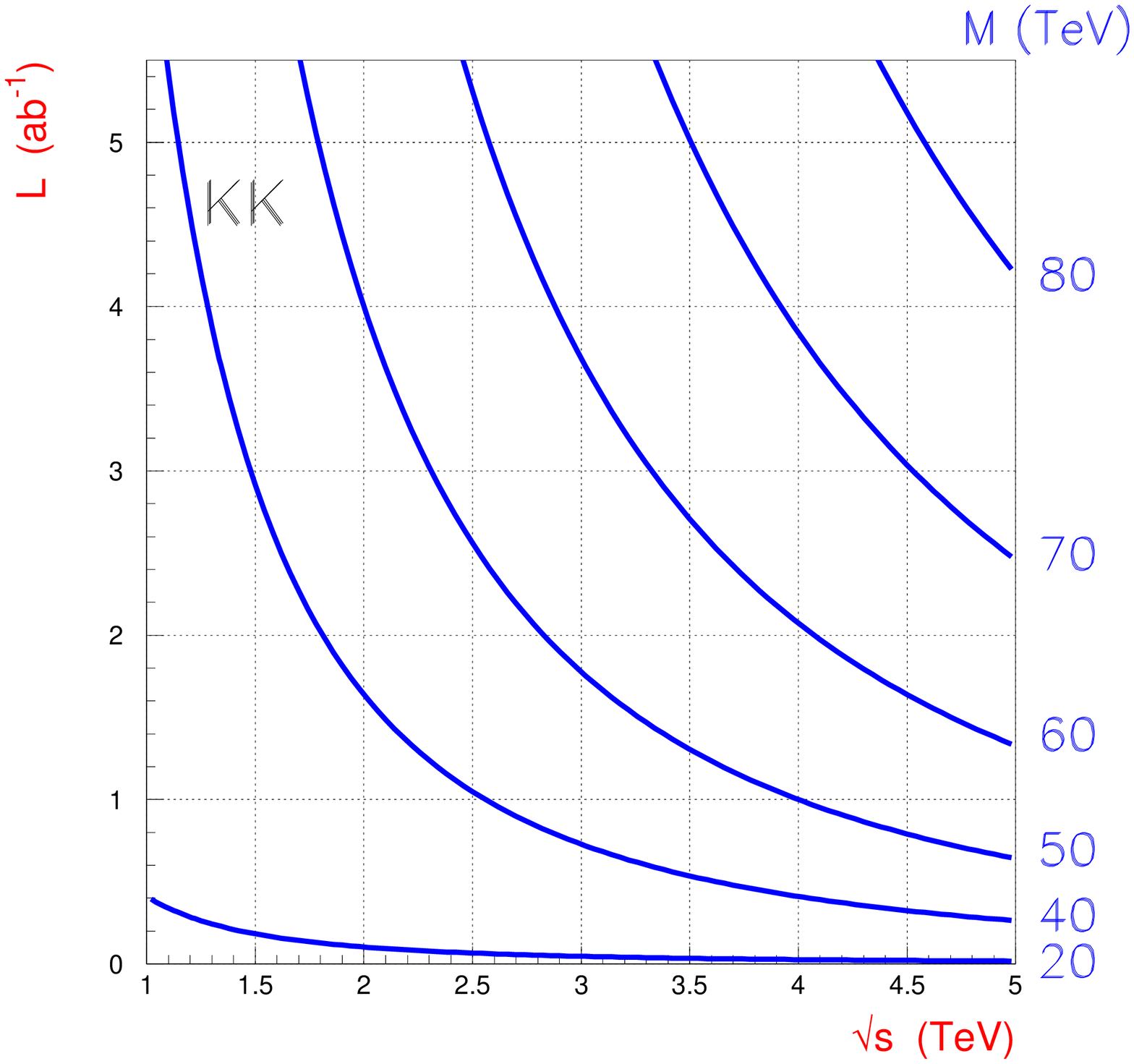}\\
\end{tabular}
\end{center}
\caption{The 95\% C.L. sensitivity contours in the $L$ vs. $\sqrt{s}$ plane
for different values of $M_{Z'}$ in the SSM model (left) 
and in the $E_6$~$\chi$ 
model (center) and for different values of the compactification scale $M$
in a five dimensional extension of the SM with fermions on the boundary 
(right).}
\label{fig:zp}
\end{figure}

The $\sigma_{f \bar f}$ and $A_{FB}^{f \bar f}$ ($f = \mu,~b$) values have been 
computed, for 1~TeV $< \sqrt{s} <$ 5~TeV, both in SM and including the corrections 
due to the presence of a $Z'$ boson with 10~TeV $< M_{Z'} <$ 40~TeV, with the couplings
predicted by the models mentioned above. This has been obtained by implementing
them in the {\sc Comphep} program. 
The relative statistical errors on the electroweak observables are obtained
by rescaling the values of Table~I for different energies and luminosities.
The sensitivity has been defined as the 
largest $Z'$ mass giving a deviation of the actual values of the observables from their 
SM predictions corresponding to a SM probability of less than 5\%.
This sensitivity has been determined, as a function of the $\sqrt{s}$ energy for 
an integrated luminosity ${\cal{L}}$ of 1~ab$^{-1}$ and 5~ab$^{-1}$, and rescaled to
other values of ${\cal{L}}$ using the formula~(2) and assuming the uncertainties to be 
dominated by statistics. Results are summarised in 
Figure~\ref{fig:zp}.  For the $\eta$ model the sensitivity is lower: for example to 
reach a sensitivity of $M_{Z'}$=20~TeV, 10~ab$^{-1}$ of data at $\sqrt{s}$=5~TeV would 
be necessary.

\section{Sensitivity to Kaluza-Klein Excitations in Theories with Extra-Dimensions}

Theories of quantum gravity have considered the existence of 
extra-dimensions for achieving the unification of gravity at 
a scale close to that of electroweak symmetry breaking. 
String theories have recently suggested that the SM could live 
on a $3+\delta$ brane with $\delta$ compactified large dimensions 
while gravity lives on the entire ten dimensional bulk.
The corresponding
models lead to new signatures for future colliders ranging from  
Kaluza-Klein (KK) excitations of the gravitons~\cite{HLZ} to KK 
excitations of the SM gauge fields with masses in the TeV 
range~\cite{antoniadis}.

Among the models with extra dimensions we have considered
the five dimensional extension of the 
SM with fermions on the boundary which 
predicts  KK excitations of the SM gauge bosons with couplings  
$\sqrt{2}$ larger than those of the SM. KK masses are given by 
$M_n=nM$, with $M$ the compactification scale of the fifth dimension.
Indirect limits on $M\sim 4$~TeV from electroweak measurements
already exist~\cite{ewlimits}. These models predict excitations of 
the $Z^0$ and of the photon which are almost degenerate in mass. 

In this analysis we have included in the cross section calculations
 only the exchange of the first 
KK excitations $Z^{(1)}$ and $\gamma^{(1)}$, neglecting the
effect of the remaining towers which give a small correction. 
The scaling law for the limit on $M$ can be obtained by considering
the interference of the two new nearly degenerate gauge 
bosons with the photon in the cross section and taking the $s<<M^2$ limit.
The result is the same as eq. (\ref{resc}). 
The analysis closely follows that for the $Z'$ boson discussed above.
In Figure~ \ref{fig:zp} we give the sensitivity contours
as a function of  $\sqrt{s}$  for different values of $M$.
We conclude that the sensitivity achievable for the compactification scale $M$ for an 
integrated luminosity of 1~ab$^{-1}$ in $e^+e^-$ collisions at 
$\sqrt{s}$ = 3~TeV - 5~TeV is of the order of 40~TeV - 60~TeV.
Results for a similar analysis, including all electro-weak observables 
are discussed in~\cite{rizzosnow}.

\section{Sensitivity to Contact Interactions}

The scenarios investigated above address specific models of New Physics beyond the SM.
\begin{table}[htbp]
 \renewcommand{\arraystretch}{1.2}
  \begin{center}
   \caption{Definition of different models of contact interaction.}
    \begin{tabular}{|c|cccccccc|}
\hline
~Model~&~LL~&~RR~&~LR~&~RL~&~VV~&~AA~&~V0~&~A0~\\
  \hline \hline
~$\eta_{\LL}$~& $\pm$1& 0    &    0   &    0  &$\pm$1 &$\pm$1 &$\pm$1 &   0 \\
~$\eta_{\RR}$~&   0   &$\pm$1&    0   &    0  &$\pm$1 &$\pm$1 &$\pm$1 &   0 \\
~$\eta_{\LR}$~&   0   & 0    &$\pm$1  &    0  &$\pm$1 &$\mp$1 &   0 &$\pm$1 \\
~$\eta_{\RL}$~&   0   & 0    &    0   &$\pm$1 &$\pm$1 &$\mp$1 &   0 &$\pm$1 \\
 \hline
    \end{tabular}
  \end{center}
\label{tab:ci}
\end{table}
\begin{figure}[htbp]
  \begin{tabular}{lr}
   \includegraphics[width=0.57\textwidth,height=0.65\textwidth]{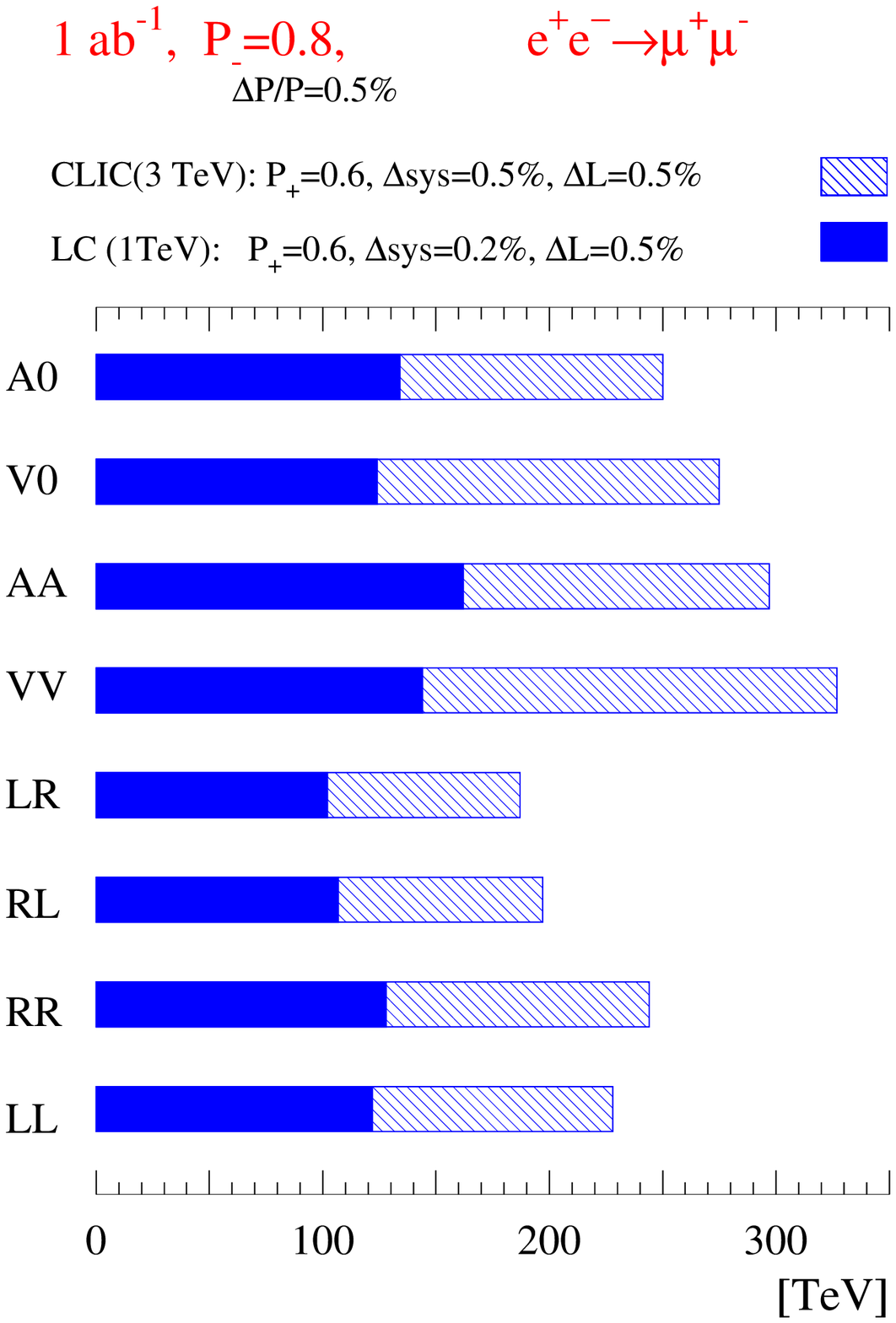}&
   \includegraphics[width=0.57\textwidth,height=0.65\textwidth]{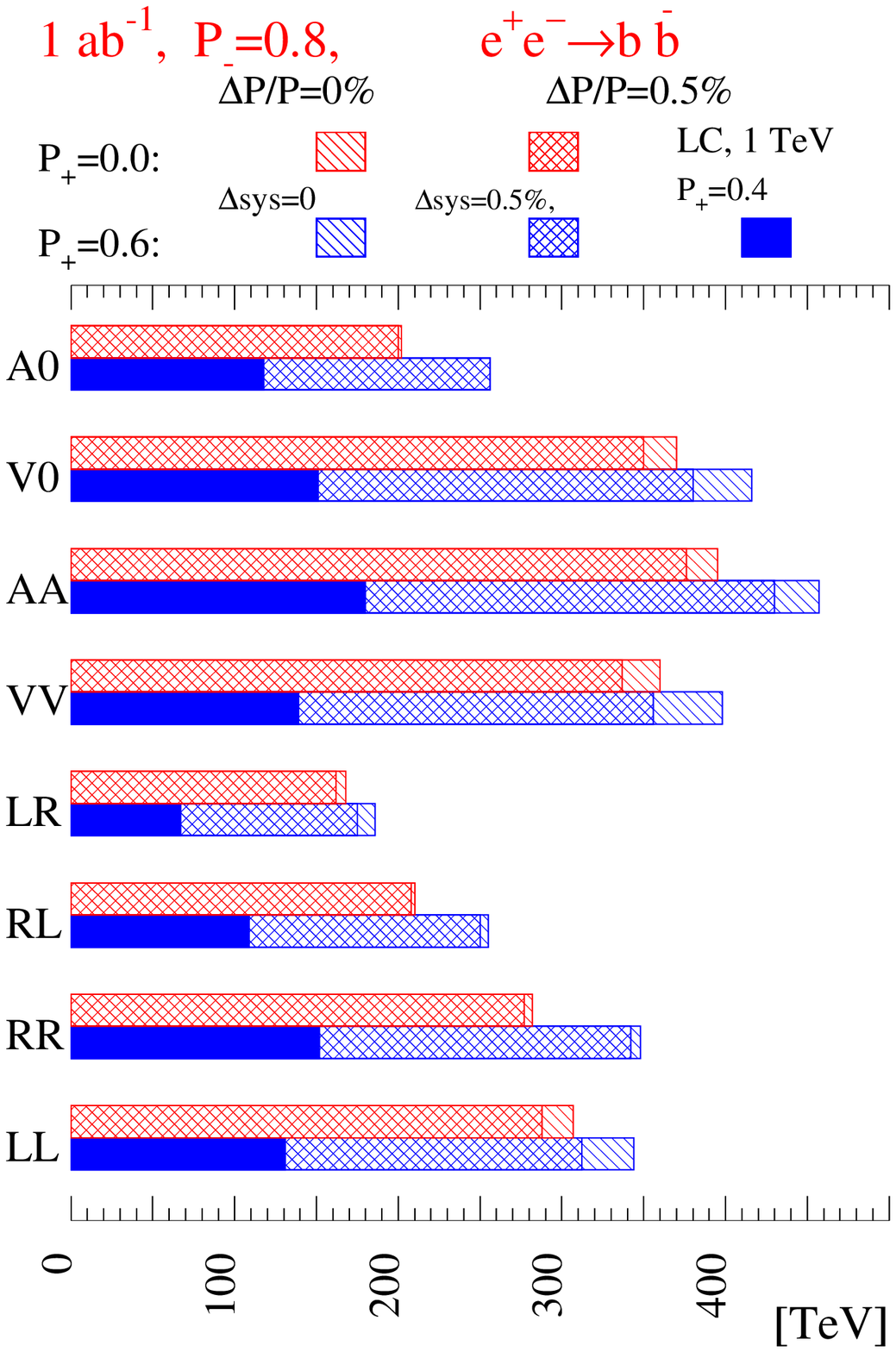}
  \end{tabular}
\caption{Limits on the scale $\Lambda$ of contact interactions in $\eemm,~\bb$
for {\sc Clic} operating at 3~TeV (dashed histogram) compared to a 1~TeV LC 
(filled histogram) for different models and the $\mu^+\mu^-$ (left) and $b \bar b$ 
(right) channels. The electron polarisations ${\cal{P}}_{-}$ is taken to be 0.8 and 
the positron ${\cal{P}}_{+}$ to be 0.6. For comparison the upper bars in the right plot
show the sensitivity achieved without positron polarisation. The influence of 
systematic uncertainties is also shown.}
\label{fig:ci}
\end{figure}

Fermion compositeness or exchange of very heavy new particles can be described in 
all generality by four-fermion contact interactions \cite{ref:ci}. These parametrise the
interactions beyond the SM by means of an effective scale, $\Lambda_{ij}$,
\begin{equation}
{\cal L}_{CI} =             \sum_{i,j = \mathrm{L,R}} \eta_{ij}
           \frac{g^2}{\Lambda^2_{ij}}
           (\bar{\mathrm{e}}_i \gamma^{\mu}\mathrm{e}_i)
           (\bar{\mathit{f}}_j \gamma^{\mu}\mathit{f}_j).
\label{ci_lagr}
\end{equation}
The strength of the interaction is set by convention as $g^2/4\pi=1$ and models
can be considered  by choosing either $|\eta_{ij}|=1$ or $|\eta_{ij}|=0$ as 
detailed in Table~II. The contact scale $\Lambda$ can be interpreted 
as effect of new particles at a mass $M_X$, $1/\Lambda^2 \propto \lambda^2 / M_X^2$.

In order to estimate the sensitivity of electro-weak observables to the contact 
interaction scale $\Lambda$, the statistical accuracies discussed in Section~II have 
been assumed for the $\mu \mu$ and $b \bar{b}$ final states. The systematics of 
the assumed 0.5\% include the contributions from model prediction uncertainties.
Results are given in terms of the lower limits on $\Lambda$ which can be excluded 
at 95\% C.L., in Figure \ref{fig:ci}. It has been verified that, for the channels 
considered in the present analysis, the bounds for the different $\Lambda_{ij}$ are 
consistent. High luminosity $e^+e^-$ collisions at 3~TeV can 
probe $\Lambda$ at scales of 200~TeV, and beyond.  
For comparison, the corresponding results expected for a LC operating at 1~TeV are also 
shown. Beam polarisation represents an important tool in these studies. 
First, it improves the sensitivity to new interactions, through the introduction of 
the left-right asymmetries $A_{LR}$ and the polarised forward-backward asymmetries 
$A_{FB}^{pol}$ in the electro-weak fits. If both beams can be polarised to 
${\cal{P}}_{-}$ and ${\cal{P}}_{+}$ respectively, the relevant parameter is the 
effective polarisation defined as 
${\cal{P}} = \displaystyle{ \frac{-{\cal{P}}_{-}+{\cal{P}}_{+}}
{1-{\cal{P}}_{-}+{\cal{P}}_{+}}}$.
In addition to the improved sensitivity, the uncertainty on the effective polarization,
can be made smaller than the error on the individual beam polarization measurements.
Secondly, in the case of a significant deviation from the SM prediction would be 
observed, $e^-$ and $e^+$ polarization is greatly beneficial to determine the nature of
the new interactions. This has been studied in details for a LC at 
0.5--1.0~TeV \cite{ref:lc-sr} and those results also apply, qualitatively, to a 
multi-TeV collider.

\section{Conclusions}

Extending the sensitivity to New Physics beyond the anticipated reach of the {\sc Lhc}, 
is a prime aim of future colliders. By accurately measuring electro-weak observables, a
LC able to achieve $e^+e^-$ collisions at and beyond 1~TeV, with high luminosity, can 
indirectly probe scales extending from tens to several hundreds TeV.

\end{document}